\documentclass[10pt,a4paper]{article}
\usepackage{amsmath}
\usepackage{graphicx}
\usepackage{epstopdf}
\usepackage{hyperref}
\usepackage{cite}

\begin{document}

\title{On the cutoff frequency of clarinet-like instruments\\
Geometrical vs acoustical regularity}
\author{E. Moers, J.Kergomard \\
Laboratoire de M\'{e}canique et d'Acoustique, CNRS UPR 7051, 31 Chemin\\
Joseph Aiguier, 13402 Marseille Cedex 20, France }
\date{}
\maketitle

\begin{abstract}
A characteristic of woodwind instruments is the cutoff frequency of their
tone-hole lattice. Benade proposed a practical definition using the
measurement of the input impedance, for which at least two frequency bands
appear. The first one is a stop band, while the second one is a pass band.
The value of this frequency, which is a global quantity, depends on the
whole geometry of the instrument, but is rather independent of the
fingering. This seems to justify the consideration of a woodwind with
several open holes as a periodic lattice. However the holes on a clarinet
are very irregular. The paper investigates the question of the acoustical
regularity: an acoustically regular lattice of tone holes is defined as a
lattice built with T-shaped cells of equal eigenfrequencies. Then the paper
discusses the possibility of division of a real lattice into cells of equal
eigenfrequencies. It is shown that it is not straightforward but possible,
explaining the apparent paradox of Benade's theory. When considering the
open holes from the input of the instrument to its output, the spacings
between holes are enlarged together with their radii: this explains the
relative constancy of the eigenfrequencies.\newline
\end{abstract}

\section{\protect\bigskip Introduction}

In his paper of 1960, Benade\cite{ben60} proposed to use the theory of
periodic media in order to analyze the effects of a row of tone-holes of
wind instruments. He was mainly interested in the length correction at the
input of a regular lattice of holes, when they are either all closed or all
open. He discovered the existence of an important frequency, the cutoff
frequency of the lattice of open holes. Below cutoff, at low frequency (in
the stop band), a wave is evanescent, i.e. exponentially decreasing, while
above cutoff (in the pass band), it can propagate. \newline

Later, in his book \cite{ben76}, he gave many details about this frequency
for real instruments and published experimental results showing the relative
independence of this frequency with respect to the fingerings (except the
fork ones) in the first register of oboes, bassoons and clarinets. In
addition he explained how this frequency is correlated to the tone-color
adjectives used by musicians to describe the overall tone of an instrument
(see \cite{ben60} p. 486).\newline
\qquad For a perfectly periodic lattice, the cutoff frequency is independent
of the fingering. For real instruments the small variation of the cutoff
with respect to the fingering suggests a great regularity of the tone-hole
lattice. However this seems to be in contradiction with the great
irregularity of the holes of a clarinet. This apparent paradox is the basis
of the motivation of the present paper, and was already noticed by Benade in
a posthumous article \cite{ben96}. Comparing 2 clarinets, one with a regular
tone hole lattice and an ordinary one, he stated that the input impedances,
measured at low level, \textquotedblleft were almost identical. This was of
course interesting and happy news, because it helped justify the use of
formal mathematical physics for a slowly varying lattice on a geometrically
quite irregular physical structure".

Benade proposed a practical method of determination of this frequency based
upon the measurement of the input impedance. Two examples will be shown in
Figs. \ref{fig:D4fijnSmooth} and \ \ref{fig:A3fijnSmooth}. In general, this
works properly even for a small number of open holes, thanks to a rather
clear distinction between the stop band and the pass band. In principle
the measured quantity is a global quantity, related to the input impedance,
which a priori depends on the whole geometry of the instrument for a given
fingering, i.e. for a given configuration of open holes. It a priori depends
therefore on the hole irregularity and the termination of the instrument. As
a consequence, the relative independence with respect to fingering is not
obvious. Notice that using the measurement of any transfer function, the
effect of a (global) cutoff frequency strictly appears only for a periodic,
infinite and lossless lattice. When losses exist, the definition is less
strict but precise. When the lattice is of finite length or/and irregular,
the definition of a global cutoff remains possible in general, as it will be
discussed in the present paper. In what follows, we will define \textit{the
global cutoff frequency as the frequency separating two frequency bands, as
viewed on the input impedance curve}. It is possible to find an analogy with
horn theory: above the global cutoff, the input impedance curve has no
resonances.

In his book, Benade also discussed the effect of irregularity. Let us cite
him (p. 449): \textquotedblleft If the lattice is irregular, theory shows
that: (1) if the first and second open-hole segments of the lattice (taken
by themselves) have widely different cutoff frequencies, the observed value
of $f_{c}$ for the composite system has an intermediate value for its cutoff
frequency; and (2) at the lower frequencies, the properties of the first
segment still dominate the implications of $f_{c}."$ We can remark that here
Benade regards the cutoff frequency as a local quantity, defined for one
segment and not for a complete lattice. We will define \textit{the local
cutoff frequency as the frequency calculated from a given cell (or segment),
corresponding to the theoretical cutoff of the periodic medium built with an
infinity of cells identical to the considered one}. In a paper on cutoff
frequencies of flutes, Wolfe and Smith \cite{wolfe} also implicitly
considered a local definition (\textquotedblleft the cutoff frequency varies
from hole to hole", see Figure 4 of the paper) and calculated it using
\textquotedblleft typical values" for the dimensions and spacing of holes,
in order to use the formula corresponding to a periodic medium. In addition
they calculated the deviation of the calculated frequencies, exhibiting a
large value for it. (For the classical flute, only the tone holes used in the
diatonic scale were included in the means. For the modern instrument, all
holes except the trill holes were included).

For a perfectly periodic lattice, i.e. a perfectly geometrically regular
lattice, there is no difficulty in defining a local cutoff,
because the lattice is divided into identical cells, one cell determining a
cutoff. For that case, local and global cutoff frequencies coincide, at
least when the lattice is long enough, and the measured global cutoff does
not depend on the fingering.

Two questions are examined in the present paper concerning a real
instrument, with irregular geometry:

\begin{enumerate}
\item Is it possible to define an \textquotedblleft acoustical
regularity\textquotedblright \footnote{%
Notice that in Ref. \cite{ben96}, Benade wrote \textquotedblleft Acoustical
regularity is a virtue", but the significance of this expression was
different: it was related to the homogeneity of input impedance for the
different fingerings.}, for which coincidence between local and global
cutoff frequencies is possible even without strict periodicity? We will
prove that the answer is positive for a lattice of holes, at least at low
frequencies, under the condition that the local cutoff frequency is uniform
and the lattice long enough. It will be explained how is possible to build
an instrument with this property. This question can be regarded as a direct
problem.

\item Concerning the inverse problem, starting from the result of Benade
that the global cutoff depends little on the fingering, and on the number of
open holes, is it possible to find an \textquotedblleft acoustical
regularity\textquotedblright\ in a real clarinet? The answer will be shown
to be positive.\ For this purpose, the delicate problem of the division of a
real clarinet into acoustically regular cells needs to be solved. The
solution is not simple, as will be shown hereafter.
\end{enumerate}

The final objective of the paper is to compare the obtained local cutoffs to
the global ones, where the local cutoffs are calculated from the geometrical
dimensions of different holes, while global cutoffs are measured for
different fingerings with one or several open holes (for this purpose, it
will be be necessary to extend the previous definition of the local cutoff).
In general the lattices are with losses and sometimes of short length,\ and
in addition they are not regular, thus the definition of the global cutoff
using the input impedance is necessarily done with a non negligible
uncertainty. In other words, there is no perfect separation between two
frequency bands. Therefore for the sought comparison, an accurate
calculation of the local cutoff is not useful and would be illusory. This
allows an approximate treatment, with simplification of the model.
Nevertheless this leads to a satisfactory comparison with experiment.

The outline of the paper is as follows: section 2 reviews the theory of
Benade, and adds a useful interpretation of the cutoff frequency as the
eigenfrequency of a cell (i.e. a segment) of the lattice. Section 3 presents
experimental results for a Yamaha Y250 clarinet for the global cutoff
frequencies measured from the input impedance, confirming Benade's results.
In addition numerical simulation exhibits the important effect of the
termination, even for a periodic lattice. Section 4 discusses the first
above-mentioned question. Section 5 tries to answer the second question: in
a first step, the tube is supposed to be cylindrical and without closed tone
holes, and in a second step some corrections are sought to this
simplification. In section 6 global and local cutoff frequencies of the
clarinet are compared and acoustical regularity is discussed.

\section{Periodic lossless lattice of holes: the cutoff frequency and its
interpretation as an eigenfrequency}

\begin{figure}[h!]
\centering\includegraphics[width=10cm]{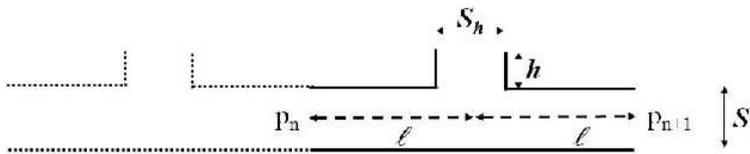}
\caption{Two $T$-shaped cells of a lattice of tone holes. A T-shaped cell
consists of the tone hole and a length $\ell $ of the main tube on both
sides of the hole. The cross section areas are $S=\protect\pi a^{2}$ and $%
S_{h}=\protect\pi b^{2}$, where $a$ and $b$ are the radii of the main tube
and the hole, respectively.}
\label{fig:geometry}
\end{figure}

Benade\cite{ben76} proposed a formula for the first cutoff frequency of a
perfectly periodic lattice of open holes, valid at low frequencies (it is
recalled below as Eq. (\ref{A5})). We remark that the corresponding
frequency is the eigenfrequency of a Helmholtz resonator built as follows:
the volume is that of a portion of the main tube with a length equal to the
spacing between two adjacent holes, and the neck is the open tone hole. In
this section we review the basic model, and explain why this remark is true,
\textit{even at higher frequencies}. The considered lattice is built with a
row of $T$-shaped cells (see Fig.\ref{fig:geometry}).

Let us consider the classical transfer-matrix description of a symmetrical
cell, relating pressures $p_{n}$ and flow rates $u_{n}$ at the extremities
(with indices $n$ and $n+1$), as follows:
\begin{equation}
\left(
\begin{array}{c}
p_{n} \\
u_{n}%
\end{array}%
\right) =\left(
\begin{array}{cc}
A(\omega ) & B(\omega ) \\
C(\omega ) & D(\omega )%
\end{array}%
\right) \left(
\begin{array}{c}
p_{n+1} \\
u_{n+1}%
\end{array}%
\right) .  \label{A1}
\end{equation}%
The transfer matrix has the following properties: coefficients $A$ and $D$
are equal, because of symmetry, and the determinant is unity, because of
reciprocity. We ignore losses, thus $A$ is real, and $B$ and $C$ imaginary
(in what follows both visco-thermal and radiation losses are ignored). In an
infinite lattice, the travelling waves are
\begin{eqnarray*}
p_{n} &=&p_{0}\exp (\pm n\Gamma ) \\
u_{n} &=&u_{0}\exp (\pm n)\text{ \ ,}
\end{eqnarray*}%
where $\Gamma $ is the propagation constant. Using Eq. (\ref{A2}) for the
two cells ($n-1$, $n$) and ($n$, $n+1$) leads to $\cosh \Gamma =A.$
Therefore at cutoff, $A=\pm 1$, $\Gamma =0$ or $\pi .$ At the cutoff
frequency, the complex amplitude of a traveling lattice wave of pressure
(respectively of flow rate), is constant from one cell to the next one, with
a factor $\pm 1$. For symmetry reasons detailed hereafter, it can be deduced
that either the flow rate or the pressure vanishes at the extremities of the
cells.

The relationship between pressure and flow rate waves is defined by the
characteristic admittance $\mathcal{Y}_{c}$ (or impedance $\mathcal{Z}_{c}=1/%
\mathcal{Y}_{c}$), given by:
\begin{equation}
\mathcal{Y}_{c}=\frac{\sinh \Gamma }{B}\text{ thus }\mathcal{Y}_{c}^{2}=%
\frac{A^{2}-1}{B^{2}}=\frac{C}{B}=\frac{C^{2}}{A^{2}-1}.\text{ }  \label{A2}
\end{equation}%
When $\Gamma $ is imaginary (and $\mathcal{Z}_{c}$ real), waves propagate,
while if $\Gamma $ is real (and $\mathcal{Z}_{c}$ imaginary), waves are
evanescent. The first case corresponds to pass bands, the second one to stop
bands.

At cutoff, $A=\pm 1$, thus $BC=0,$ i.e. \textit{either }$B$\textit{\ or }$C$%
\textit{\ vanishes}. If $C$ vanishes$,$ $\mathcal{Y}_{c}$ vanishes too. The
flow rate being proportional to $\mathcal{Y}_{c}$ for the waves in the two
directions, it is zero for any value of $n$, for both an infinite or a
finite lattice. Therefore the pressure wave is constant:%
\begin{equation*}
p_{n}=Ap_{n+1}=\pm p_{n+1}.
\end{equation*}
If $p_{n}=p_{n+1}$, the pressure field is symmetrical, while if $%
p_{n}=-p_{n+1}$, the pressure field is antisymmetrical. The dual situation,
reversing the roles of pressure and flow rate, occurs if $B=0$. Finally the
cutoff frequencies are the eigenfrequencies of a cell with either Neumann ($%
u_{n}=0$, if $C=0$) or Dirichlet ($p_{n}=0$, if $B=0)$ conditions at the
extremities.

The next question is the distinction between the first eigenfrequencies
satisfying the termination conditions. At low frequencies, for a cell of a
tone-hole lattice, the coefficient $A$ is larger than unity , therefore the
waves are evanescent and the first cutoff occurs for $A=+1,$ i.e. when
either the pressure field or the flow rate field is symmetrical.\ As it is
well known, a Helmholtz resonator has an eigenfrequency very low when its
volume is closed, therefore the first cutoff frequency of the lattice, which
is the subject of the present paper, corresponds to the Neumann boundary
conditions, with a symmetrical pressure field. This is true even if the
wavelength at cutoff is not larger than the dimensions of a cell. In Ref.
\cite{chaigne08} expressions are given for the four types of cutoff
frequencies (see also Appendix \ref{app2} of the present paper), with a
comparison of the first four ones, confirming the fact that the lowest
cutoff is that of the Helmholtz resonance of a T-shaped cell.\ It
corresponds to the condition $\mathcal{Y}_{c}=0$, or $C=0$, and this is in
accordance with a mathematical analysis done by Benade (in his Eq. 8, the
cutoff is obtained when the denominator vanishes \cite{ben60}). To our
knowledge, this interpretation is new.

It can be concluded that considering the transfer matrix or the equivalent
circuit of a tone-hole (see Refs \cite{keefe82, dubos}), the element
corresponding to the antisymmetrical field, which is a series impedance
denoted $Z_{a},$ does not appear in the expression of the first cutoff
frequency (this is a rigorous result, without any approximation).\
Nevertheless ignoring this series impedance is not valid at any frequency.
In Appendix \ref{app2}, it is shown how this series impedance could be taken
into account in order to generalize the present approach, but for the
purpose of the paper, simplified models are sufficient and we ignore it.

Now, if the height of the hole chimney is assumed to be much shorter than
the wavelength, and if losses are ignored, the effect of the tone hole is
reduced to a shunt acoustic mass, denoted $m_{h}$. The coefficients for
transfer matrices of a $T-$shaped cell are given by standard acoustic
theory:
\begin{eqnarray}
A &=&D=1-2\sin ^{2}k\ell +jYz_{c}\sin k\ell \cos k\ell  \label{B2} \\
B &=&z_{c}\left[ 2j\sin k\ell \cos k\ell -Yz_{c}\sin ^{2}k\ell \right]
\notag \\
C &=&z_{c}^{-1}\left[ 2j\sin k\ell \cos k\ell +Yz_{c}\cos ^{2}k\ell \right]
\text{ ,}  \notag
\end{eqnarray}%
where $Y=(j\omega m_{h})^{-1}$ is the shunt admittance of the hole and $%
z_{c}=\rho c/S$, the characteristic impedance of the main tube. $\rho $ is
the air density, $c$ the sound speed, $S=\pi a^{2}$ the cross-section area
of the tube, assumed to be cylindrical, $2\ell $ the spacing between two
holes, $\omega $ the angular frequency, and $k=\omega /c$ the wavenumber.

The equation satisfied by the cutoff frequency is given by $C=0$:
\begin{equation}
j\frac{\rho c}{S}\cot k\ell =2j\omega m_{h}\text{ .}  \label{A3}
\end{equation}%
The left-side member is the impedance on both sides of the tone hole,
deduced from the Neumann boundary condition, while the right-side member is
twice the shunt impedance of the tone hole. The mass $m_{h}$ is the sum of
the mass of the planar mode in the hole and of the masses corresponding to
the radiation impedance into surrounding space and into the main \ tube. It
is approximately equal to:
\begin{equation}
m_{h}=\rho h_{t}/S_{h}\text{, with }h_{t}\simeq h+1.6b,  \label{A4}
\end{equation}%
where $h$ is the height of the hole, and $S_{h}=\pi b^{2}$ the cross section
area of the hole. Notice that $h_{t}$ is denoted $t_{e}$ by Benade.

Solving Eq. (\ref{A3}) when $k\ell <<1$ leads to the result (Ref. \cite%
{ben76}):
\begin{equation}
f_{c}=\frac{c}{2\pi }\frac{1}{\ell }\frac{1}{\sqrt{2m_{h}/m}}=\frac{c}{2\pi }%
\frac{b}{a}\frac{1}{\sqrt{2\ell h_{t}}}.  \label{A5}
\end{equation}
where $m=\rho \ell /S$ is the acoustic mass of the portion of the main tube
of length $\ell $ (notice that in Eq. (\ref{A5}) the compressibility of air
in the main tube appears, via the acoustic compliance $C_{a}=\ell S/\rho
c^{2},$ but its inertia does not). The exact value of $h_{t}$ depends on
several parameters, such as the undercutting of the hole or the existence of
a key pad, but this is not critical for the present study, especially
because the cutoff depends on the square root of this mass.

A better approximation for the solution of Eq. (\ref{A3}) is the following
(see Refs \cite{kergo81, keefe90}):
\begin{equation}
f_{c}=\frac{c}{2\pi }\frac{1}{\ell }\frac{1}{\sqrt{2m_{h}/m+1/3}}=\frac{c}{%
2\pi \ell \sqrt{2(a/b)^{2}h_{t}/\ell +1/3}}.  \label{A6}
\end{equation}%
It is obtained by expanding $\cot k\ell $ to the next order in $k^{2}\ell
^{2}$. This gives a condition of validity of Eq. (\ref{A5}):
\begin{equation}
\frac{b^{2}}{a^{2}}<<6\frac{h_{t}}{\ell }\text{ \ \ or \ }\frac{m}{m_{h}}<<6%
\text{ ,}  \label{A7}
\end{equation}%
If $k_{c}=2\pi f_{c}/c$, this condition is equivalent to: $k_{c}^{2}\ell
^{2}<<1$ (the half spacing between two holes is much smaller than the cutoff
wavelength, i.e. the elements of the system are lumped). For a clarinet,
typical values of the cutoff frequency and length $\ell $ are $1500$ Hz and $%
10$ mm, thus $k_{c}\ell \sim 0.3$, and the condition is satisfied.\bigskip

The dimensions and locations of the holes are given in Appendix \ref{app1}
(Table 1).\ Table 2 of the same appendix indicates the first opened holes
for the different fingerings.

\section{Determination of global cutoff frequencies}

Benade\cite{ben76} proposed a simple method to measure the \textquotedblleft
global\textquotedblright\ cutoff frequency of a given instrument,
considering the curve of the input impedance modulus. The first two
frequency bands generally appear rather clearly: the first one with high and
regular peaks, the second one with smaller and irregular peaks. Benade
defined the cutoff as the boundary between the two bands. As stated in the
introduction, in principle this method is perfect for a perfectly periodic
lattice (i.e. regular, lossless and infinite). What happens for a real
lattice is discussed hereafter.

The explanation given by Benade (p.\ 434) is based upon the strong radiation
of the holes above cutoff. This can be more detailed: below cutoff, in the
stop band (waves are evanescent), the effective length of the tube is very
close to the tube cut at the first open hole, thus the frequency interval
between resonances is large. Moreover boundary layer losses (which are
preponderant in this range) are small, thus the impedance peaks are high.

On the contrary, above cutoff (in the pass band), the effective tube is
divided into two portions. The first one is without open holes, while the
second one is the open-hole lattice. The phase velocity and characteristic
impedance are different in the two portions, thus the impedance peaks are
irregular. Moreover boundary-layer losses exist over a large length, and
several open holes radiate efficiently. Therefore the peaks are lower than
in the stop band. The efficient radiation is due not only to the number of
holes radiating, but also to the external interaction between the holes, as
shown in Ref. \cite{belgrade}.

Benade\cite{ben76} measured modern and baroque instruments, and found that
baroque instruments have lower cutoff frequencies than modern ones. The
explanation seems to be evident thanks to the analysis of the previous
section: first of all, the holes of baroque instruments are generally
narrower than those of modern instruments. In addition baroque instruments
are basically diatonic instruments while roughly speaking the basis of
modern instruments is more chromatic. Therefore spacings between open holes
are larger for baroque instruments than for modern ones. These two facts
with the interpretation of the cutoff frequency as the eigenfrequency of a
cell viewed as a resonator explain the differences in cutoff frequencies. A
consequence is the slightly wider compass of modern instruments, even if it
is possible to play notes with frequencies higher than the cutoff (obviously
a complementary explanation for the wider compass of modern instruments is
the addition of new holes). Otherwise the question of the influence of the
cutoff frequency on the sound spectrum has been discussed rather rarely, but
Benade and Kouzoupis can be cited \cite{ben88}, as well as Ref. \cite{wolfe}
for the flutes. This question is out of the scope of the present paper.
Finally the question of the influence of cutoff on directivity of woodwinds
has been treated in Refs. \cite{ben60, ben80, kergo81, belgrade}.

\subsection{Measurement results}

\begin{figure}[h]
\centering
\includegraphics[width=10cm]{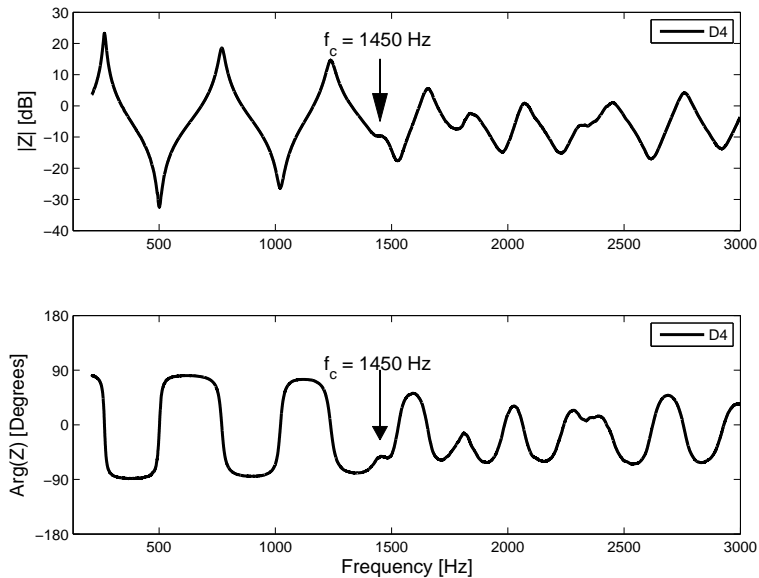}
\caption{Measured impedance curve for note D4. The cutoff $f_{c}$ is found
to be 1450. The scale for the impedance modulus is the logarithmic one: $%
20\log (|Z|/z_{c}).$}
\label{fig:D4fijnSmooth}
\end{figure}
\begin{figure}[h]
\centering
\includegraphics[width=10cm]{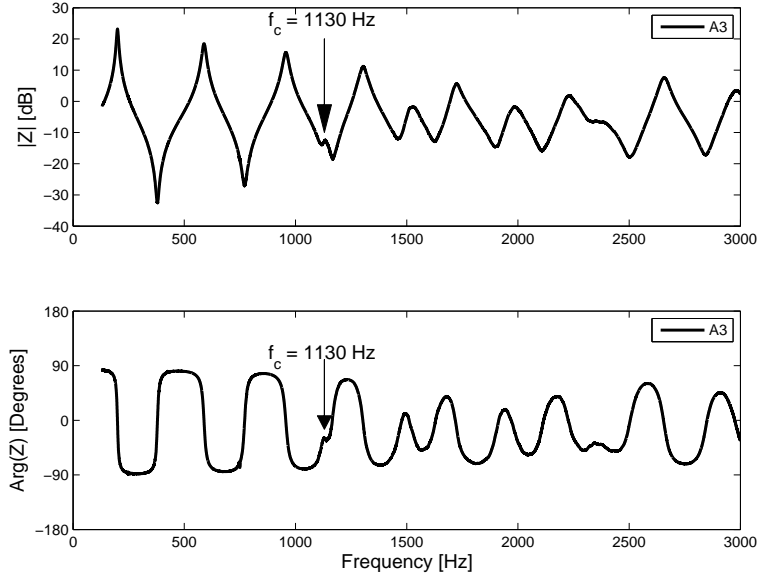}
\caption{Measured impedance curve for note A3.\ The cutoff $f_{c}$ is found
to be 1130 Hz. The scale for the impedance modulus is the logarithmic one: $%
20\log (|Z|/z_{c}).$}
\label{fig:A3fijnSmooth}
\end{figure}
\qquad

Benade deduced the cutoff frequency from the measured modulus of the input
impedance on a linear scale. Actually it is easier in practice to use either
the modulus of the impedance on a logarithmic scale or its argument. This is
often better because a slope inversion\ clearly appears for almost every
fingering, even when only a small number of holes is open. This leads to a
definition of the global cutoff with a precision in practice better than
1\%. Nevertheless, as stated in the introduction, this does not mean that
the separation of the two frequency bands is precise, the definition being
somewhat arbitrary. A further discussion is given in \cite{FA}.

Figures \ref{fig:D4fijnSmooth} and \ \ref{fig:A3fijnSmooth}\ \ show two
examples of measured input impedance curves for the notes D4 and A3. For the
note D4 (262\ Hz), the global cutoff is found to be 1450 Hz, while for the
note A3 it is 1130 Hz.\ For all the notes of the first register, the
measurement of the cutoff frequency is easy, even when only one hole is open
(note F3): Benade and Kouzoupis\cite{ben88} explained this fact by the
effect of the bell, \textquotedblleft which serves as a more or less
surrogate for an open-hole lattice\textquotedblright\ (sect VIID, see also
Ref. \cite{ben96}).

An interesting result is that the cutoff does not vary very much (a
variation of 20\% is much smaller than the variations of geometrical
dimensions), even for this kind of notes, as it will be seen on Figure \ref%
{fig:MeasFc}, which shows the results for the different notes of the
clarinet studied (Yamaha Y250). The results are within the range of results
obtained by Benade, who measured several different instruments. Notice that
Benade gave results for the first register, for the same notes as those we
have studied, except the notes for which the first open tone hole is
provided with what we call a \textquotedblleft closed key\textquotedblright
, i.e. a key open for this note only (see Appendix \ref{app1} Table \ref%
{tab:notes}). Otherwise, as expected, the cutoff frequencies for the second
register, when the register hole is open, are very close to the cutoffs for
the first register, for the corresponding fingering.

For the first register, two groups of notes can be observed, above and below
B4, around 1450 Hz and 1150 Hz, respectively. We will see in section \ref%
{SIR} that this difference is not related to regularity, but it is due to
the termination effect.
\begin{figure}[h!]
\centering
\includegraphics[width=10cm]{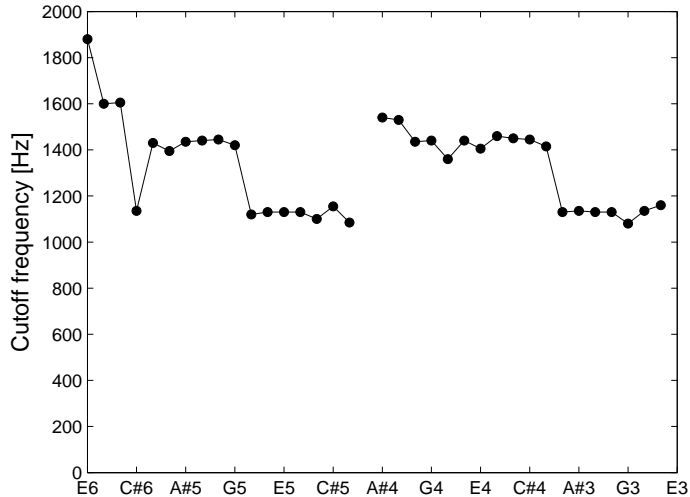}
\caption{Cutoff frequencies measured for the Yamaha Y250 clarinet. The graph
covers all fingerings, and for the first register exhibits a great
similarity with the results of Benade, the scale being chosen to be similar.}
\label{fig:MeasFc}
\end{figure}

The device used for the impedance measurement is based upon the measurement
of acoustic pressures in two cavities separated by a flow rate source,
marketed by CTTM \cite{Z}.

\subsection{Simulation results\label{SIR}}

In order to get more insight into the problem, some simple simulations using
transfer matrices like Eq. (\ref{B2}) have been carried out.\ No series
impedances are taken into account (see Eqs.\ (\ref{B2})), but boundary layer
losses and radiation, given by standard theory, are considered. Radiation
takes external interaction into account, via a global admittance matrix (see
Ref. \cite{belgrade}). A simplified shape of the bell is used. Some open
holes are considered through a unique, equivalent tone hole, as explained
hereafter in section \ref{analysis}, resulting in a lattice with 11 open
holes. The (global) cutoff frequencies are deduced using the input impedance
curve.

\begin{figure}[h!]
\centering
\includegraphics[width=12cm]{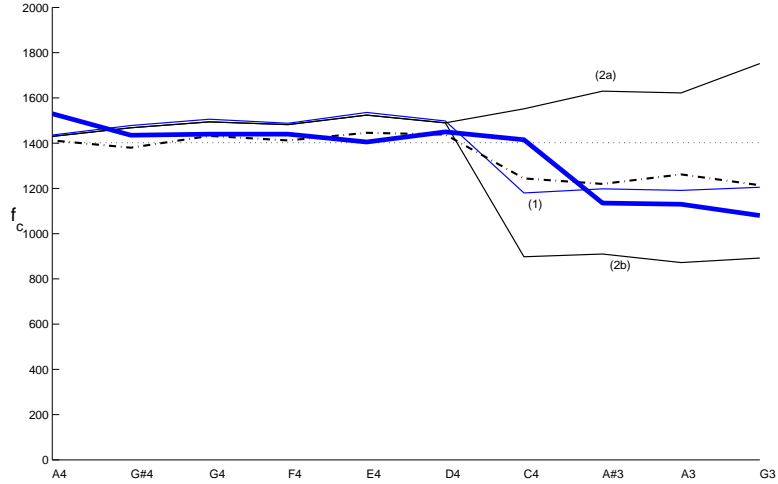}
\caption{Measured and calculated global cutoff frequencies. Some fingerings
are not considered, see explanation in table \protect\ref{tab:notes} in
Appendix \protect\ref{app1}; moreover the two lowest notes, E3 and F3, are
not considered as well, only notes with at least two open holes being
considered. Solid, thick line: experimental results (also shown in the
previous figure). Dashdot line: numerical result of the simplified model
with a bell (see section \protect\ref{asym}). Solid, thin lines: numerical
results for a purely periodic lattice: (1) with a bell, (2) with a
cylindrical tube replacing the bell (for low notes, two possible values of
the cutoff are shown, denoted 2a and 2b). The thin, dotted line indicates
the theoretical cutoff of the periodic lattice. }
\label{fig:MeasFcAllReg}
\end{figure}

Fig.\ref{fig:MeasFcAllReg} shows an interesting qualitative agreement
between this model and experiment, sufficient for our purpose. The
discrepancy is less than 11\%, this value occurring for the note A$\sharp$3.
This fact can be related to the location of the limit between the two groups
of values, near to A$\sharp$3.

The most interesting result is the comparison between the values for the
geometrical data of the studied clarinet and those for a perfectly periodic
lattice, having 11 open holes with constant spacing $2\ell =0.0341$m and
theoretical cutoff $1450$ Hz, the radius being $7.5$ mm. The total length is
the same. The common value of the reduced masses, $m_{h}/\rho =340$ m$^{-1}$%
, is deduced from Eq. (\ref{A5}). Notice that the exact value (Eq. (\ref{A3}%
)) of the theoretical cutoff is $1402$ Hz; the approximation (\ref{A6}),
giving also $1402$ Hz, is excellent.

The main features are the following:

\begin{itemize}
\item The differences between the results for the purely periodic lattice
and the simplified model of the irregular, real one are less than 5\%; this
can be seen as a first indication of the existence of an acoustic regularity;

\item The existence of two groups of values with a limit for A3 is roughly
similar for the two lattices;

\item Above A3 (first group), when many holes are open, the global cutoff is
higher than the theoretical value (1402 Hz), some values being higher than
1500 Hz, the average being 6\% higher than the theoretical one. Even for the
highest note, when all the holes are open, the global cutoff is 2\% higher
than the theoretical one.

\item The existence of two groups is due to \textit{the effect of
termination only}. This can be checked by replacing the bell by a
cylindrical tube of same length and input radius. The values corresponding
to the lowest notes (2$^{nd}$ group) are strongly modified. The above cited
sentence by Benade and Kouzoupis is probably true, because for the lowest
notes, the determination of the global cutoff is uncertain. For the
cylindrical termination, irregular peaks are found around 900 Hz, but the
shape of the impedance curves differs strongly from the typical curves shown
in Figures \ref{fig:D4fijnSmooth} and \ \ref{fig:A3fijnSmooth}.
\end{itemize}

\begin{figure}[h]
\centering
\includegraphics[width=10cm]{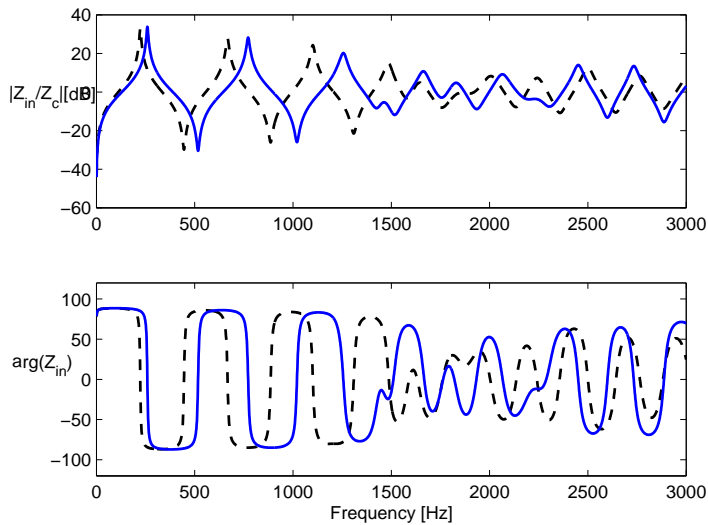}
\caption{Calculated input impedance curve for the fingering D4.\ Solid line:
simplified model of the clarinet. Dashdot line: periodic lattice of the same
length and termination. .}
\label{figcalcul}
\end{figure}

Figure \ref{figcalcul} shows a comparison of theoretical results for
two kinds of lattices: the one of the considered clarinet, and the
above mentioned periodic lattice\footnote{
  between the first resonance frequencies comes from the difference in
  length of the tubes upstream from the first open tone hole. The
  spacing between the first tone holes is significantly smaller than
  the mean spacing, used for the periodic lattice. This will be seen
  in Fig. \ref{A11}.}
bell). .

For this particular case, it appears that the practical definition of the
global cutoff is easier for the real (irregular) lattice than for periodic
one with the same termination. It is a confirmation that the definition of
the global cutoff is not always easy in practice (For an infinite lattice
without losses, the argument should exhibit a discontinuity between the two
bands).\

\section{Construction of an acoustically regular lattice \label{AGR}}

This section investigates if acoustically regular lattices can exist. It is
known since Anderson\cite{anderson} that in a onedimensional medium, the
effect of an infinite number of random irregularities is the suppression of
pass bands, and therefore of cutoff frequencies. Obviously this asymptotic
property cannot be observed on musical instruments, because of their limited
length. Moreover the theorem of F\"{u}rstenberg\cite{furstenberg} concerning
the product of random matrices indicates that some exceptions to the
Anderson's result can exist. In Ref. \cite{depollier}, it is shown that the
product of matrices having the same characteristic impedance $\mathcal{Z}%
_{c} $ is such an exception. As a matter of fact, for that case:
\begin{eqnarray}
\underset{i}{\overset{N}{\Pi }}\left(
\begin{array}{cc}
\cosh \Gamma _{i} & \mathcal{Z}_{c}\sinh \Gamma _{i} \\
\mathcal{Z}_{c}^{-1}\sinh \Gamma _{i} & \cosh \Gamma _{i}%
\end{array}%
\right) &=&\left(
\begin{array}{cc}
\cosh \sigma & \mathcal{Z}_{c}\sinh \sigma \\
\mathcal{Z}_{c}^{-1}\sinh \sigma & \cosh \sigma%
\end{array}%
\right)  \label{B1} \\
\text{where \ }\sigma &=&\underset{i}{\overset{N}{\Sigma }}\Gamma _{i}.
\notag
\end{eqnarray}

The behavior is similar to this of the regular medium with the same total
propagation constant $\sigma =n\Gamma $. As a consequence, if a lattice is
built with irregular cells having the same characteristic impedance \emph{at
every frequency}, its behavior is the same as the behavior of a perfectly
periodic medium. If this situation can exist for wind instruments, acoustic
regularity can exist without geometrical regularity. In particular stop and
pass bands can exist: when $\sigma $ is imaginary (and $\mathcal{Z}_{c}$
real), waves propagate, while if $\sigma $ is real (and $\mathcal{Z}_{c}$
imaginary), waves are evanescent. This is now investigated.

The cutoff of a $T$-shaped cell is given by $A=1$ or $C=0$ (see Eq. (\ref{A3}%
)). In order to exhibit the value of the cutoff wavenumber $k_{c}$, using
Eq. (\ref{A3}) for $\omega =\omega _{c}$ and $k=$ $k_{c}$, the mass $m_{h}$
can be eliminated and the admittance $Y=(j\omega m_{h})^{-1}$ is rewritten
as follows:
\begin{equation*}
Y=-2jz_{c}^{-1}\frac{k_{c}}{k}\tan k_{c}\ell \text{ .}
\end{equation*}%
Thus, using the definition of the transfer matrix (Eq. (\ref{B2})):
\begin{equation}
\mathcal{Z}_{c}^{2}=\frac{B}{C}=z_{c}^{2}\frac{1+\frac{k_{c}}{k}\tan
k_{c}\ell \,\tan k\ell }{1-\frac{k_{c}}{k}\tan k_{c}\ell \,\cot k\ell }.
\label{B3}
\end{equation}%
The characteristic impedance is written with respect to two quantities, the
half-length $\ell $ of a cell and the cutoff wavenumber $k_{c}$, the latter
parameter replacing the hole mass. If a lattice is built with cells of
identical $k_{c}$, the characteristic impedance can be identical (i.e.
independent of $\ell $) at low frequencies if both $k_{c}\ell $ and $k\ell $
are small quantities:
\begin{equation}
\mathcal{Z}_{c}=\frac{B}{C}=z_{c}^{2}\frac{1+O(k_{c}^{2}\ell ^{2})}{%
1-k_{c}^{2}/k^{2}}.  \label{B4}
\end{equation}%
The propagation constant is given by $\sinh ^{2}\Gamma =BC$ with
\begin{equation}
BC=-4\sin ^{2}k\ell \cos ^{2}k\ell \,\;(1+k_{c}k^{-1}\tan k_{c}\ell \tan
k\ell )(1+k_{c}k^{-1}\tan k_{c}\ell \,/\tan k\ell )  \label{B411}
\end{equation}%
thus with the same approximation,
\begin{equation}
\Gamma =2j\varphi \ell \text{ \ \ with \ }\varphi =k\sqrt{1-\frac{k_{c}^{2}}{%
k^{2}}}\text{.}  \label{B41}
\end{equation}%
$\varphi $ is an equivalent wavenumber. Therefore, in the frequency range
where the cell dimensions are smaller than wavelength (and consequently
where Eq. (\ref{A5}) is valid), it is possible to build an acoustically
regular lattice, provided that the cutoff frequency of the cells is a
constant. The length of the cells can be arbitrarily chosen but cannot be
too long, and for each cell the value of the hole acoustic mass is deduced
from the chosen value of the cutoff frequency. Notice that starting from the
input of the tube, the distance between holes can either increase or
decrease, and consequently the hole masses can decrease or increase,
respectively (e.g. the hole radii increase or decrease). The property of the
considered lattice is identical to that of a purely periodic one, therefore
the global cutoff is the same as the local cutoff of the cells. This is the
answer to the first question in the introduction.

Eqs. (\ref{B4}) and (\ref{B41}) suggest an analogy with the problem of an
exponential horn: this is discussed in Ref. \cite{chaigne08}. Studying the
``horn function'' of a given bell as Benade did, is equivalent to studying
the local cutoff frequency of the bell, which has strong variation for
instance for a trumpet bell\cite{ben73}.

This result concerning acoustical regularity will be more complicated if the
antisymmetrical (negative) masses are taken into account in the calculation,
but the answer remains positive (see Appendix \ref{app2}). Other
complications of the model are possible if they can be compatible with the
basic model, based upon the association of an acoustic mass with an acoustic
compliance. Fortunately this is in general the case at low frequencies.

\section{The inverse problem; analysis of an irregular lattice}

\subsection{Statement of the problem}

We will now analyze the lattice of a real instrument. Solving the inverse
problem, i.e. dividing a given lattice into cells having the same cutoff
frequency, is not an easy task, because the solution either is not unique or
does not exist, as explained hereafter. Obviously a first requirement for a
method of division is to re-obtain the initial division when considering a
lattice built as explained in section \ref{AGR}, instead of a real one.

We first assume the clarinet to be a purely cylindrical tube with holes of
different sizes. As it is known, the radius of the main tube does not vary
very much for a clarinet. We will see how it is possible to take into
account conicity of some portions as corrections. Another hypothesis has
been made: in a first step the closed holes have no influence, and their
effect is taken as another kind of correction.

For the present purpose, we consider 14 open holes.\ In three cases (holes
number 8 and 9, 14 and 15, 18 and 19, see Appendix \ref{app1}, table \ref%
{tab:notes}), two closely spaced holes are opened simultaneously in order to
get a given note: according to the basic hypothesis of long wavelength, we
choose to replace them by a single hole, with a mass equivalent to the two
masses in parallel, and located at the middle of the interval of the two
holes.\ It remains 11 equivalent holes, therefore 11 cells. All the other
holes are closed.

Three methods of analysis have been investigated, the first two being
based upon different choices of division of the bore into cells, the
third trying to extend the definition of a local cutoff frequency
without any division.

\subsection{Methods of analysis and results \label{analysis}}

\subsubsection{Division into symmetrical cells with varying eigenfrequency}

A first method is implemented to divide the tone-hole lattice into $T$%
-shaped, symmetrical cells. It is not possible to fix a common value for the
cutoff frequency, because doing that all the cell lengths become fixed (they
are deduced from the values of the hole masses and of the eigenfrequencies),
and the cells either will have overlap or do not re-build the complete
lattice.

Thus for the chosen method no value for the cutoff is a priori fixed. An
initial parameter is arbitrarily chosen, i.e. for instance the half length
of the first (upper) cell in the lattice, denoted $\ell _{1}$. Using
iteration, this implies the length of each cell, therefore the whole
division of the lattice. The cutoff frequencies of each cell can be deduced.
They are a priori different, and depend on the chosen value for $\ell _{1}$%
.\ The ratio $R_{c}$ of the cutoff highest value to the lowest one is
calculated, and the final choice of the parameter $\ell _{1}$ is found by
searching for the minimum value of this ratio, which appears very clearly.

The method is first tested on an ideal (acoustically regular) lattice of 11
tone-holes, as defined in the previous section. As expected, the minimum
ratio $R_{c}$ is unity, for a length $\ell _{1}$ equal to the half length of
the first cell. The method is therefore capable of dividing correctly a
lattice built to be acoustically regular.

On the contrary, when applied to the lattice of a clarinet, no division has
been found, because negative spacings between holes appear. This is probably
due to the close location of the two first upper holes. When ignoring these
two holes, a result is found, but the value of the minimum ratio is 2.8:
this value is high, while the other approaches, as it will be discussed
hereafter, give much smaller values, i.e. a much more uniform value for the
cutoff frequencies. Therefore this method has been abandoned \cite{moers}.

\subsubsection{Division into asymmetrical cells with constant eigenfrequency
\label{asym}}

For the second method, a more general model of acoustically regular lattices
is investigated, with one degree of freedom more. Asymmetrical cells are
considered: the hole is not necessarily located at the middle of the cell.
As a matter of fact, the location of the neck of a Helmholtz resonator has a
small influence on its eigenfrequency. The essential elements are an
acoustic mass and an acoustic compliance, related to the volume. At low
frequencies, it is therefore possible to modify a (symmetrical) $T$-shaped
cell by moving the input and output by the same length, $\delta ,$ without
changing the transfer matrix (the condition being $\delta <<\lambda $, if $%
\lambda $ is the wavelength). The accuracy of the model based upon this
division is a priori similar to that of the division into symmetrical cells
(this will be discussed more precisely when comparison with measured global
cutoff frequencies will be presented).

The division into asymmetrical cells makes available a supplementary
parameter. It is possible to set a constant value for the
eigenfrequency of the different cells. From the knowledge of the
eigenfrequency and the hole mass, the lengths $\ell _{ln}+\ell _{rn}$
of the cells are obtained. The starting point is an initial value for
the length $\ell _{l1}$ to the left of the first open
hole. Therefore, from the knowledge of $\ell _{ln}$, the length
$\ell _{rn}$ on the right of the hole is deduced, then, from the spacing
between two holes, the length $\ell _{ln+1}$, etc.
\begin{figure}[h]
\centering
\includegraphics[width=10cm]{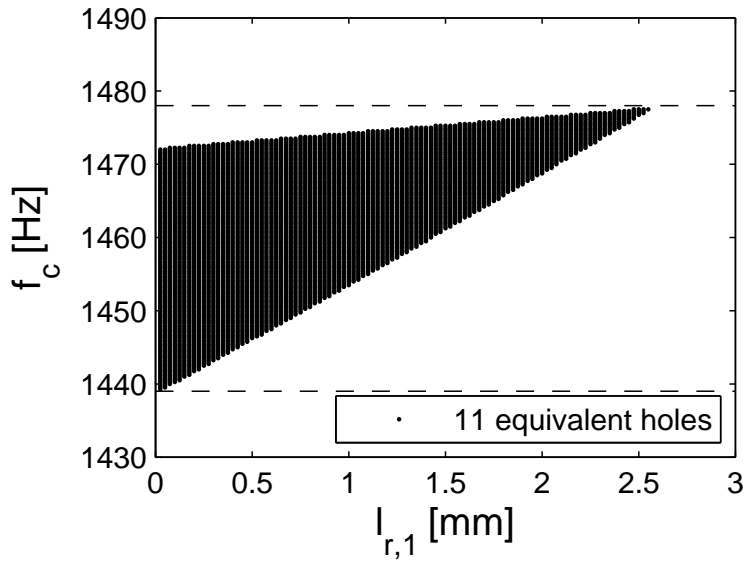}
\caption{2$^{nd}$ method: Division of a lattice with 11 equivalent holes
into asymmetrical cells. On the horizontal axis is the semi-cell length to
the right $\ell _{r1}$. The region where a solution (i.e. a possible common
eigenfrequency) is found is dashed and the horizontal dotted lines limit the
frequency range where a correct partition exists.}
\label{fig:GraphMethod2Holes11Klein}
\end{figure}
Depending on the choice of the initial length $\ell _{l1},$ a more or less
wide range of possible eigenfrequencies is found. The result is that a
solution exists for $f_{c}\in \lbrack 1439,1478]$ Hz, as shown in Fig.\ref%
{fig:GraphMethod2Holes11Klein} (there is a continuum of solutions).

\begin{figure}[h]
\centering
\includegraphics[width=10cm]{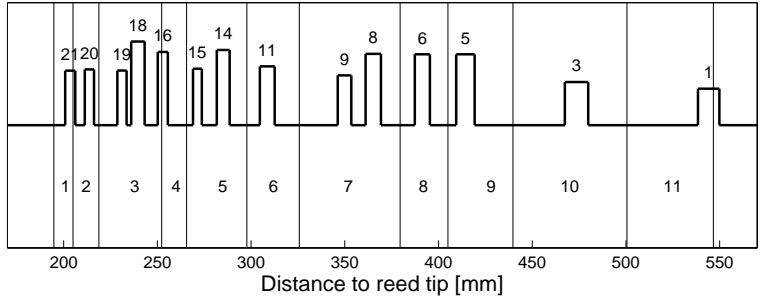}
\caption{2$^{nd}$ method: Illustration of the partition into 11 asymmetrical
cells for $f_{c}=1470$ Hz and $l_{r,1}=1.50$mm. The numbers below are the
cell numbers from the input of the instrument; while the numbers on the top
are the hole numbers, as defined in Appendix \protect\ref{app1}. For three
cells, two holes have been replaced by an equivalent hole.}
\label{A11}
\end{figure}

Looking at the partition itself, graphed in Fig.\ref{A11} for $f_{c}=1470$
Hz and $l_{r,1}=1.50$mm, it is observed that some cells are very
asymmetrical, the borders being located very close to the middle of a hole
(and even within the hole opening). This seems to be curious and unreal, but
formally the transfer matrix of the whole lattice is identical to this of an
acoustically regular one, and we will see in the final discussion that the
results are interesting. The acoustical regularity can be far from the
geometrical one!

The surprise can diminish if we accept that the found lattice is equivalent
to the symmetrized lattice with the same holes and cells, but with holes at
the center of the cells. We checked that the main discrepancies between the
asymmetrical lattice and the symmetrized one occur at frequencies much
higher than cutoff. Nevertheless the relative error is non negligible at
very low frequencies, and this is intuitive: makers know that the shift of
the first open tone hole modifies the first resonance frequencies. Further
analytical analysis confirms that the expected error due to the moving of
the holes is larger at low frequencies than around cutoff. Finally we notice
that this symmetrized lattice cannot be found by the first method, which is
not flexible enough.

\subsubsection{A possible extension of the definition of a local cutoff
frequency\label{local}}

A third method of analysis is not based upon any possible division. If two
adjacent, symmetrical $T$-shaped cells \textit{have the same eigenfrequency }%
with different lengths $\ell _{1}$ and $\ell _{2}$ and different hole masses
$m_{h1}$and $m_{h2},$ Eq. (\ref{A5}) leads to:
\begin{equation}
\ell _{1}m_{h1}=\ell _{2}m_{h2}=\frac{\rho }{2Sk_{c}^{2}}\text{ ,}
\label{C1}
\end{equation}%
therefore the spacing $d=\ell _{1}+\ell _{2}$ between the holes satisfies:
\begin{equation}
d=\frac{1}{k_{c}^{2}}\frac{\rho }{2S}\left( \frac{1}{m_{h_{1}}}+\frac{1}{%
m_{h2}}\right)
\end{equation}%
thus
\begin{equation}
f_{c}=\frac{c}{2\pi }\sqrt{\frac{\rho }{2Sd}\left( \frac{1}{m_{h_{1}}}+\frac{%
1}{m_{h2}}\right) }.  \label{C3}
\end{equation}%
This quantity can be calculated without a division of the lattice for every
pair of tone holes. It is the eigenfrequency of a resonator of length $d$,
with two necks corresponding to the holes with a cross section divided by 2.
If it is a constant over the length of a lattice, the lattice is
acoustically regular. If it is not constant, its variation can be regarded
as a measure of irregularity. We can define the frequency given by Eq. (\ref%
{C3}) as a local cutoff frequency, depending in a direct way on the
dimensions (masses) of two adjacent holes and their distance. This extend
the definition given in the introduction, and the two definitions are
coherent for either geometrically or acoustically regular lattices.

\begin{figure}[h!]
\centering
\includegraphics[width=12cm]{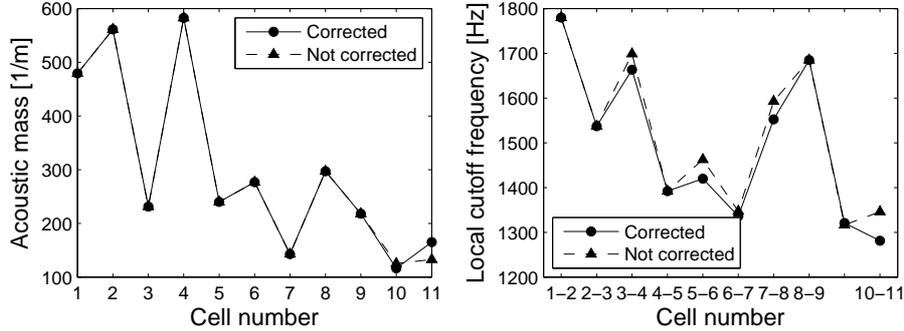}
\caption{3$^{rd}$ method. Left figure: Acoustic mass divided by the density $%
m_{h}/\protect\rho $, for 11 equivalent tone holes. Right figure: Local
cutoff frequency for a pair of adjacent holes (Eq. (\protect\ref{C3})). For
both quantities, the uncorrected and corrected models (see section \protect
\ref{corr}) are compared. }
\label{fig:Method3MAholesCorrComp}
\end{figure}

Fig.\ref{fig:Method3MAholesCorrComp} shows the results of the third method.
An important feature is the difference in variation for the acoustic masses
and the local cutoff frequencies. The maximum variation for the square root
of the masses is 2.24 while the maximum variation for the cutoff frequencies
is 1.38. It can be concluded that the choice of the spacing between the tone
holes allows a significant compensation for the variation of the hole
acoustic masses, and a certain acoustical regularity exists. This is
confirmed by a rough inspection of the holes of a real instrument: the holes
appear to be larger as well as their spacing from the input to the bell. Is
this effect directly sought by the makers? It is far from evident, and to
our minds this remains an open question.

\subsection{Results with corrections \label{corr}}

The objective of the present work is to analyze a real lattice in terms of
acoustical regularity, and implies use of a simple model. Actually many
details have an influence on the local cutoff frequencies, but the order of
magnitude of this influence remains small.\ In order to validate this idea,
two types of corrections have been studied:

\begin{itemize}
\item the effect of closed holes. At low frequencies, this effect is due to
the air compressibility, and is proportional to the volume of the cavity of
the closed hole. The volume of the portion of the main tube involved in Eq. (%
\ref{A5}) is therefore modified, in accordance with the lumped elements
hypothesis. Fig.\ref{fig:Method3MAholesCorrComp} shows that the correction
for the cutoff is small. The cutoff frequency is lowered by a typical amount
of 25 Hz.

\item the effect of the cross-section variation of the main tube. We are
interested in the enlargement of the portion of the tube where open holes
are present.\ This portion is not long. The choice is to describe this
enlargement as the insertion of a change in conicity, just below hole n$%
{{}^\circ}%
3$ (see Appendix \ref{app1}), by using results explained e.g. in Ref. \cite%
{ben88c}. This change in conicity is represented by a supplementary shunt
mass $m_{cone}$, which is rather high, i.e. equivalent to a narrow open
hole. It is given by the following formula:
\begin{equation*}
m_{cone}=\rho \frac{x}{S}
\end{equation*}%
where $x$ is the length of the missing part of the cone, equal to $287$mm$.$
The mass is inserted at $28.4$mm from hole n$%
{{}^\circ}%
3.$ In addition, the masses of holes n$%
{{}^\circ}%
1$ and n$%
{{}^\circ}%
2$ need to be multiplied by the ratios $S_{1}/S$ and $S_{2}/S$, where $S_{i}$
is the cross section of the main tube at the location of hole n$%
{{}^\circ}%
i.$ Again the correction of the results appears to be very small (see Fig.%
\ref{fig:Method3MAholesCorrComp}).
\end{itemize}

Concerning the corrections of the results of the 2$^{nd}$ method of analysis
(division into asymmetrical cells), the effect is small as well. The range
of possible common eigenfrequencies becomes slightly narrower and lower.

\section{Global and local cutoff frequencies}

Using the results obtained in the previous section, it is possible to
compare the theoretical eigenfrequencies and the measured, global cutoff
frequencies. It is reasonable to think that the use of a simplified model
does not modify the discussion written hereafter, because the corrections
are very small. Nevertheless the results take the two kinds of corrections
into account.

\begin{figure}[h!]
\centering
\includegraphics[width=10cm]{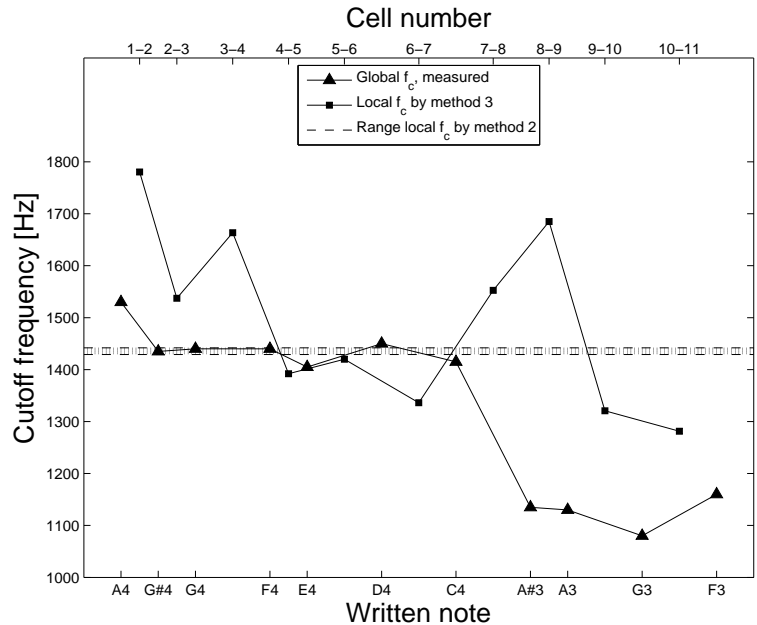}
\caption{Comparison between experimentally determined global $f_{c}$ and
local $f_{c}$ obtained using method 2 and 3 for the corrected model. The
values of $f_{c}$ for method 3 are plotted between the tones, because they
are based on the acoustic mass of two subsequent equivalent tone holes. The
dashed region shows the frequency range for the local $f_{c}$ obtained by
the 2$^{nd}$ method.}
\label{fig:CompM3ExpFcCorr}
\end{figure}

Fig.\ref{fig:CompM3ExpFcCorr} shows the results of: i) the measurements of
the global cutoffs; ii) the calculation of the possible constant
eigenfrequencies using division into asymmetrical cells (2$^{nd}$ method \S %
\ref{asym}); iii) the calculation\ of the local cutoff frequencies (3$^{rd}$
method, \S \ref{local}). Notice that there are two different types of axes
for the abscissa: for the theoretical results, the numbers correspond to the
cell numbers, while for the experimental ones, the results depend on the
fingerings.

The most evident feature is the satisfactory result of the division into
asymmetrical cells. The constant eigenfrequency obtained by this method
coincides very well with the measured global cutoff for many fingerings. We
think that this is a validation of both the definition of the acoustical
regularity and the method of division into asymmetrical cells. Nevertheless we
notice that this excellent agreement is the consequence of a slight
overestimation of the two results: for the theoretical result, as mentioned
earlier, the frequency is higher than the true one, because of the use of
approximation (\ref{A5}); for the experimental result, it has been seen that
the practical measurement of the global cutoff overestimates the true cutoff
as well. Otherwise a discrepancy exists for four lower notes. The
explanation has been given earlier, in section \ref{SIR}, when studying the
effect of the termination.

\bigskip Concerning the 3$^{rd}$ method, it gives an order of magnitude of
the local cutoffs, but they are in general 15\% higher than the global
measured values, at least at the ends of the considered register. The
tendency of the variation looks rather similar, except for fingerings around
the note A3 (we have no interpretation of this fact). If for a given
fingering the global cutoff was determined by the value of the cell
corresponding to the first open hole, the two frequencies should coincide
(notice however that the results of this method concern a pair of tone
holes). It is not the case, but the fact that the tendency is similar
(except for some notes) is in accordance with the hypothesis that the global
frequencies are determined by the local frequencies of the first cells.

We finally remark that the 3$^{rd}$ method seems to be interesting because
of its simplicity, no division being needed. \ For sake of simplicity, the
improvement of Eq. (\ref{C3}) by taking into account the correction term of
Eq. (\ref{A6}) is not discussed here. The correction depends on the length
of the cell, but the irregularity of the results shown in Fig. \ref%
{fig:CompM3ExpFcCorr} is not significantly reduced.

\section{\protect\bigskip Conclusion}

\bigskip A theoretical definition of acoustical regularity is possible, at
least at low frequencies, and can be applied to a clarinet. \textit{An
important degree of acoustical regularity is found on a clarinet, despite
the rather great geometrical irregularity. }This explains why the measured
cutoff frequencies do not vary very much, except when the termination effect
occurs. Results for another type of clarinet probably should be rather
similar (see especially Ref.\cite{wolfeweb1}).

The acoustical regularity is limited to wavelengths large compared to the
inter-hole spacing. A consequence is that if higher cutoff frequencies
exist, e.g. limiting a new stop band, the previous analysis of regularity
cannot be expected to be relevant for these frequencies. A second stop band
seems to exist on both experimental and numerical results between roughly
2500 Hz and 3000 Hz, the second cutoff being different from one fingering to
another one. This frequency band was noticed by Wolfe \cite{wolfeweb2}. The
understanding of the existence of this band can be a subject for future
investigation: what is sure is that it is far from the theoretical next
cutoff frequency of the periodic lattice studied in section \ref{SIR}. This
frequency, called the Bragg frequency in physics, is the half-wavelength
eigenfrequency of a cell without holes, i.e. $c/(4\ell ),$ equal to 5000 Hz,
see e.g. Ref. \cite{kergo81}).

The division of a real lattice as an acoustically regular lattice is not an
easy task. Thanks to an extension of the definition of such a lattice, the 2$%
^{nd}$ method gives a satisfactory division. Notice that when Benade wrote
about the cutoff frequency of segments (in the sentences cited in the
introduction), he did not explain how this frequency is defined, i.e. how
the two first segments are divided.

The present paper does not present a classical comparison between experiment
and theory: a precise comparison between theory and experiment could be
sought, especially by taking into account the effect of key pads, the
antisymmetrical effects of tone holes, or the precise geometry of the bell.
However the agreement between measured global cutoff frequencies and the
theoretical cutoff of the acoustically regular lattice built from the real
geometry is very good. The limitation to long wavelength is not problematic:
this was not evident, because though the spacing between holes is small
compared to the cutoff wavelength, the total length of the lattice is not
small at all.

The 3$^{rd}$ method is very simple and gives correct orders of magnitude:
the concept is probably rather close to that in the mind of Benade. When
qualitatively looking at the location and sizes of the tone holes (see Fig.%
\ref{fig:GraphMethod2Holes11Klein}), the correlation between larger spacings
between holes and wider holes roughly appears, and this is confirmed by the
approach of the 3$^{rd}$ method.

It remains to understand the origin of this correlation. Why do the makers
provide an increase of the spacing between holes together with an increase
of their radius? Probably it is related to the search for correct tuning,
because when a hole is moved downstream, it needs to be enlarged for a given
tuning. If this is true, why do the makers enlarge the holes far from the reed?\
Is it related to radiation efficiency, or to nonlinear effects, or to the
possibilities of the fingers and the keys? Another topic for future
investigation is a more complete understanding of the effect of the value of
the cutoff frequency on the tone color.

In this paper the question of the bell of the clarinet has not been studied
in a precise way. Its effect is known to ensure a correct tuning of the second
register (see refs \cite{neder, neder2, debut}). An idea could be that for
tone-color purposes, the shape of the bell, which is nearly catenoidal, is
sought to be equivalent to a continuation of the open tone-hole lattice.
According to Ref. \cite{neder}, the cutoff of the (infinite) catenoidal horn
should be $f=mc/2\pi $, where $m$ is the expansion parameter, found to be $%
1/0.085$ m$^{-1}.$ This would lead to a value of $636$ Hz, much lower than
the cutoff of the tone-hole lattice.\ Therefore this simple idea is not
satisfactory, and anyway it ignores the finite size effect of periodic media.

Finally we remark that an extension of this study to other cylindrical
woodwinds, such as the flute, and even to conical instruments, such as
saxophones or oboes, is possible without great complexity.

\begin{center}
APPENDICES
\end{center}

\appendix{\label{app}}

\section{Geometry of the clarinet studied\label{app1}}

The dimensions and locations of the holes of the clarinet Yamaha Y250 are
given in Table 1. Table 2 indicates the first opened holes for the different
fingerings.
\begin{table}[h]
\caption{Numerical data for the complete set of tone holes. Holes are
numbered for decreasing distance to the tip of the reed. Hole 24 is the
register hole. Holes 22 and 23 are not used for basic fingerings (see Table
\protect\ref{tab:notes}), therefore they have not been considered in the
study. The uncertainties of the measurements are $\pm 0.02$mm for the
tone-hole radius, $\pm 0.04$mm for the tone-hole height, and $\pm 0.005$mm
for the tube radius. }
\label{tab:lattice}\centering
\begin{tabular}{c|c|c|c|c}
\hline\hline
$no.$ & $x$ [mm] & Hole radius $b$ [mm] & Bore radius $a$ [mm] & Hole height
$h$ [mm] \\ \hline
24 & 156.50 & 1.50 & 7.53 & 12.50 \\
23 & 167.33 & 2.46 & 7.50 & 6.90 \\
22 & 193.98 & 2.88 & 7.46 & 6.87 \\
21 & 203.51 & 2.70 & 7.46 & 6.66 \\
20 & 213.74 & 2.47 & 7.46 & 6.81 \\
19 & 231.14 & 2.47 & 7.46 & 6.68 \\
18 & 239.66 & 3.58 & 7.47 & 10.23 \\
17 & 241.78 & 2.48 & 7.47 & 6.89 \\
16 & 252.90 & 2.69 & 7.47 & 8.95 \\
15 & 271.38 & 2.36 & 7.47 & 6.91 \\
14 & 285.12 & 3.44 & 7.47 & 9.20 \\
13 & 287.43 & 2.75 & 7.47 & 6.60 \\
12 & 289.95 & 2.83 & 7.47 & 6.67 \\
11 & 308.62 & 3.94 & 7.47 & 7.19 \\
10 & 318.88 & 2.61 & 7.47 & 6.66 \\
9 & 349.86 & 3.65 & 7.49 & 6.08 \\
8 & 365.19 & 4.13 & 7.51 & 8.72 \\
7 & 370.36 & 3.80 & 7.52 & 6.99 \\
6 & 391.39 & 4.02 & 7.54 & 8.66 \\
5 & 414.34 & 4.91 & 7.55 & 8.66 \\
4 & 446.63 & 5.25 & 7.53 & 5.13 \\
3 & 473.60 & 6.22 & 7.52 & 5.27 \\
2 & 505.78 & 5.70 & 7.67 & 4.63 \\
1 & 544.10 & 5.71 & 8.40 & 4.46 \\ \hline\hline
\end{tabular}%
\end{table}

\begin{table}[h]
\caption{Most common notes of a B$^{\flat}$ clarinet (written
  pitches). The third column gives the target frequency of a well
  tuned note when it is normally tempered. The fourth column gives the
  hole number of the first open hole(s) in the lattice. When no number
  is listed, this note is not considered in this study and does not
  belong to the \textquotedblleft normal" set of fingerings. Excluding
  a note from this set means that for its corresponding fingering, it
  is necessary to open a tone hole that keeps closed for the other
  fingerings. }
\label{tab:notes}\centering
\begin{tabular}{r|l|c|l}
\hline\hline
& Note & $f_p$ $[\text{Hz}]$ & First opened tone-hole number \\ \hline
Chalumeau & E3 & 147 & all closed \\
& F3 & 156 & 1 \\
& F$\sharp$3 & 165 & - \\
& G3 & 175 & 3 \\
& G$\sharp$3 & 185 & - \\
& A3 & 196 & 5 \\
& A$\sharp$3 & 208 & 6 \\
& B3 & 220 & - \\
& C4 & 233 & 8+9 \\
& C$\sharp$4 & 247 & - \\
& D4 & 262 & 11 \\
& D$\sharp$4 & 277 & - \\
& E4 & 294 & 14+15 \\
& F4 & 311 & 16 \\
& F$\sharp$4 & 330 & - \\ \hline
Throat & G4 & 349 & 18+19 \\
& G$\sharp$4 & 370 & 20 \\
& A4 & 392 & 21 \\
& A$\sharp$4 & 415 & - \\ \hline
Clarinet & B4 & 440 & all closed \\
& C5 & 466 & 1 \\
& C$\sharp$5 & 494 & - \\
& D5 & 523 & 3 \\
& D$\sharp$5 & 554 & - \\
& E5 & 587 & 5 \\
& F5 & 622 & 6 \\
& F$\sharp$5 & 659 & - \\
& G5 & 698 & 8+9 \\
& G$\sharp$5 & 740 & - \\
& A5 & 784 & 11 \\
& A$\sharp$5 & 831 & - \\
& B5 & 880 & 14+15 \\
& C6 & 932 & 16 \\ \hline\hline
\end{tabular}%
\end{table}
\clearpage

\section{Use of a more complete model\label{app2}}

We have ignored the negative acoustic masses corresponding to the
antisymmetrical field in the holes. If the series impedance $Z_{a}=j\omega
m_{a}$ is taken into account together with the shunt admittance $Y=(j\omega
m_{h})^{-1}$, the transfer matrix of a hole is written as follows (see e.g
Fig.3 of Ref. \cite{dubos}):
\begin{equation*}
\frac{1}{1-Z_{a}Y/4}\left(
\begin{array}{cc}
1+Z_{a}Y/4 & Z_{a} \\
Y & 1+Z_{a}Y/4%
\end{array}%
\right) .
\end{equation*}%
Multiplying this matrix on both sides by the transfer matrix of a segment of
cylindrical tube of length $\ell $ leads to the coefficients of the matrix
of the T-shaped cell:
\begin{eqnarray*}
B &=&\left( \cos k\ell +j\frac{Y}{2}z_{c}\sin k\ell \right) \left( Z_{a}\cos
k\ell +2jz_{c}\sin k\ell \right) \left( 1-Z_{a}Y/4\right) ^{-1} \\
C &=&\left( Y\cos k\ell +2jz_{c}^{-1}\sin k\ell \right) \left( \cos k\ell +j%
\frac{Z_{a}}{2}z_{c}^{-1}\sin k\ell \right) \left( 1-Z_{a}Y/4\right) ^{-1} .
\end{eqnarray*}%
This result exhibits the four types of cutoff frequencies, the lowest one
corresponding to the vanishing of the first factor of the coefficient $C$.
As expected, all of them depend on either the series impedance or the shunt
admittance, for reasons of symmetry. Therefore the exact value of the
cutoffs are simpler than those obtained after approximations, as has been
done in Ref. \cite{keefe90}.

In order to study the acoustical regularity, Eq. (\ref{B3}) is transformed
into:
\begin{equation*}
\mathcal{Z}_{c}=\frac{B}{C}=z_{c}^{2}\,\frac{1+\frac{k_{c}}{k}\tan k_{c}\ell
\,\tan k\ell }{1-\frac{k_{c}}{k}\tan k_{c}\ell \,\cot k\ell }\;\,\frac{1-j%
\frac{Z_{a}}{2}\cot k\ell }{1+j\frac{Z_{a}}{2}\tan k\ell }.
\end{equation*}%
At low frequencies, the characteristic impedance given by Eq. (\ref{B3}) is
multiplied by the factor $(1+m_{a}/2m)^{1/2}$. As a consequence, it is
possible to ensure an improved acoustical regularity, as follows: in order
to obtain a constant characteristic impedance at every frequency, both the
cutoff $f_{c}$ and the ratio $(1+m_{a}/2m)/S^{2}$ can be chosen to be
equal.\ Therefore, for a given length $\ell $, there are two equations for
the two parameters of the holes (height $h$ and radius $b$).

However the ratio $m_{a}/2m\simeq -0.18b^{3}/(a^{2}\ell $) is in general
close to $0.01$ or $0.02$, thus it is not important to take this term into
account in the present study, because a high precision is not needed.

\section*{Acknowledgments}

We thank Didier Ferrand and Alain Busso for their help for the experiments,
and Ren\'{e} Causs\'{e}, Jean-Pierre Dalmont, Douglas Keefe, Philippe
Guillemain, Avraham Hirschberg, Franck Lalo\"{e}, Kees Nederveen, P.-A.
Taillard and Joe Wolfe for useful discussions. We thank also the referees and Murray Campbell
for their help in improving the manuscript.

\end{document}